\documentclass[fleqn,twoside]{article}
\usepackage[headings]{espcrc2}
\usepackage[english, USenglish]{babel}
\usepackage{amsmath}
\usepackage{amsfonts}
\usepackage{latexsym}
\usepackage{bm}
\usepackage[bookmarksnumbered=true,breaklinks=true]{hyperref}
\usepackage{epsfig}
\usepackage{a4wide}
\usepackage{espcrc2}
\newcommand{\hepth}[1]{\href{http://www.arXiv.org/abs/hep-th/#1}{\tt hep-th/#1}}
\newcommand{\hepph}[1]{\href{http://www.arXiv.org/abs/hep-ph/#1}{\tt hep-ph/#1}}

\readRCS
$Id: espcrc2.tex,v 1.2 2004/02/24 11:22:11 spepping Exp $
\usepackage{graphicx}
\usepackage[figuresright]{rotating}

\title{\begin{flushright}{\small CERN-PH-TH/2007-024}\end{flushright}
\bf{Topological Amplitudes and Physical Couplings in String Theory}\thanks{Proceedings for the Cargese Summer School 2006, Lectures by I. Antoniadis}}

\author{I.~Antoniadis\address[CERN]{Department of Physics, CERN -- Theory Division\\CH-1211 Geneva 23, Switzerland} 
        \thanks{\tt ignatios.antoniadis@cern.ch},
        S.~Hohenegger\addressmark\thanks{\tt stefan.hohenegger@cern.ch}}
       
\runtitle{\bf{Topological Amplitudes and Physical Couplings in String Theory}}
\runauthor{I.~Antoniadis and S.~Hohenegger}

\begin{document}
\begin{abstract}
In these lectures, we review the main properties of the topological theory obtained by twisting the $\mathcal{N}=2$ two-dimensional superconformal algebra, associated to supersymmetric string compactifications. In particular, we describe a set of physical quantities in string theory that are computed by topological amplitudes. These are in general higher-dimensional $F$-terms in the low energy effective supergravity, or fermion masses after supersymmetry breaking. We discuss $\mathcal{N}=2$ compactifications of type II strings, $\mathcal{N}=1$ compactifications of heterotic and type I strings, as well as $\mathcal{N}=4$ string vacua. Particular emphasis is put on alternative string dual representations allowing calculability, and on the generalization of $\mathcal{N}=2$ holomorphicity and its anomaly.
\end{abstract}

\maketitle

\tableofcontents
\newpage
\section{$\mathcal{N}=2$ Superconformal Algebra and the 2d Topological Twist}
{\it The two-dimensional (2d) $\mathcal{N}=2$ superconformal algebra is introduced and the topological twisting of the theory is explained. The topological partition function is defined and the special role of Calabi-Yau-threefold compactifications is shown. The holomorphic anomaly is described, which leads to a recursion relation for the antiholomorphic moduli dependence of the partition function.} 
\subsection{$\mathcal{N}=2$ Superconformal Algebra}\label{sec:N2SCFT}
Four-dimensional (4d) string compactifications with $\mathcal{N}=1$ space-time supersymmetry ($\mathcal{N}=1+1$ for type II strings) are described by an underlying 2d $\mathcal{N}=2$ superconformal field theory (SCFT)~\cite{Banks:1988yz}.
The $\mathcal{N}=2$ superconformal algebra contains the energy momentum tensor $T$, two (conjugate) supercharges $G^\pm$ and a $U(1)$ current $J$:\footnote{We will restrict our considerations to the left-moving sector only, since the algebra of right-movers follows trivially in precisely the same way. In our conventions, right-movers are marked by a bar: $(\bar{T},\bar{G}^{\pm},\bar{J})$.}
\begin{eqnarray}
T(z)T(w)&=&\frac{c}{2(z-w)^4}+\nonumber\\
&&+\frac{2T(w)}{(z-w)^2}+\frac{\partial_wT(w)}{z-w},\\
T(z)G^{\pm}(w)&=&\!\!\frac{3G^\pm(w)}{2(z-w)^2}+\frac{\partial_wG^\pm(w)}{z-w},\\
T(z)J(w)&=&\frac{J(w)}{(z-w)^2}+\frac{\partial_wJ(w)}{z-w},\\
G^+(z)G^-(w)&=&\frac{2c}{3(z-w)^3}+\frac{2J(w)}{(z-w)^2}+\nonumber\\
&&+\frac{2T(w)+\partial_wJ(w)}{z-w},\label{modalg}\\
J(z)G^\pm(w)&=&\pm\frac{G^\pm(w)}{z-w},\label{anomaly}\\
J(z)J(w)&=&\frac{c}{3(z-w)^2}.
\end{eqnarray}
The constant $c=\frac{3\hat{c}}{2}$ is the central charge, while the conformal dimension and $U(1)$ charge of the operators are displayed in the following table:
\begin{center}
\begin{tabular}{ccc}\hline
\textbf{operator} & \textbf{conf. weight} & \textbf{$U(1)$}\\ \hline
$T$  & $2$ & $0$\\
$G^\pm$ & $3/2$ & $\pm1$\\
$J$ & 1 & 0\\\hline
\end{tabular}
\end{center}
\subsection{The Topological Twist and the Partition Function}\label{sec:toptwist}
Following \cite{Witten:1992fb,Bershadsky:1993cx,Cecotti:1991me,Lerche:1989uy} the \textbf{topological field theory} is obtained by twisting the energy-momentum tensor subtracting (half of) the derivative of the $U(1)$ current:
\begin{align}
T\to T-\frac{1}{2}\partial J,
\label{toptwist}
\end{align}
such that the new central charge vanishes and (\ref{modalg}) of the superconformal algebra is changed to
\begin{align}
G^+(z)G^-(w)\sim \frac{2J(w)}{(z-w)^2}+\frac{2T(w)}{z-w}+\ldots
\end{align}
The twist also results in shifting the conformal dimension of all operators by half their $U(1)$ charge:
\begin{align}
h\to h-\frac{q}{2},\hspace{1cm}\text{with}\ \left\{\begin{array}{l}h=\text{conf. weight}\\ q= U(1)\text{ charge}\end{array}\right.\label{topshift}
\end{align}
Thus, the new conformal weights are:
\begin{center}
\begin{tabular}{ccc}\hline
\textbf{operator} & \textbf{conf. weight} & \textbf{$U(1)$}\\ \hline
$T$   & $2$ & $0$\\
$G^+$ & $1$ & $1$ \\
$G^-$ & $2$ & $-1$\\
$J$ & $1$ & 0\\\hline
\end{tabular}
\end{center}

One can now identify $G^+$ with the BRST operator
\begin{align}
Q_{\text{BRST}}=\oint G^+,
\end{align}
which immediately shows that the energy momentum tensor becomes BRST exact:
\begin{align}
\{Q_{\text{BRST}},G^-\}= \oint G^+G^-=T.
\end{align}
Thus, $G^-$ can be identified with the reparametrization anti-ghost which can be sewed with the $(3g-3)$ Beltrami differentials of a genus $g$ Riemann surface to define the integration measure over its moduli space $\mathcal{M}_g$. It is then straight-forward to write down the expression for the topological partition function
\begin{align}
F_g=\int_{\mathcal{M}_g}\langle |\prod_{a=1}^{3g-3}G^-(\mu_a)|^2\rangle.\label{toppartition}
\end{align}
One may think naively that this expression vanishes by charge conservation. However, in the twisted theory, the anomaly (\ref{anomaly}) of the $U(1)$ current provides a background charge of $\hat{c}(g-1)$ on genus $g$. 
In the case of $\hat{c}=3$, the integration measure of the topological partition function (\ref{toppartition}) alone is sufficient to cancel this deficit and the partition function can therefore be non-vanishing without additional operator insertions. The `critical value' $\hat{c}=3$ is reached for compactifications on Calabi-Yau-threefolds ($\text{CY}_3$). An interesting property of these compactifications is the notion of \textbf{special geometry} which, although very important in many applications, will not be studied in this review.

From (\ref{topshift}), one immediately encounters that the fields, which fulfill (before the twist)
\begin{align}
h=\pm\frac{1}{2}q, \hspace{1cm}\text{with } q=\pm 1,
\end{align}
play a special role and are called chiral (anti-chiral) primary fields $\phi$ ($\bar{\phi}$). They satisfy $[G^+,\phi]=0$. It follows that the chiral primaries, which obtain conformal weight $h=0$ after the topological twist, are the cohomology states of the BRST operator:
\begin{align}
Q_{\text{BRST}}\phi=\oint G^+\phi=0,
\end{align}
and span the physical Hilbert space. In the same way, the anti-chiral primaries acquire conformal weight $h=1$ after the twist and are BRST exact. Hence, they are unphysical states and should decouple from the topological partition function $F_g$. This is however true only up to an anomaly that we discuss below.
\subsection{Holomorphic Anomaly Equation}
According to the previous section, a perturbation of the action of the form 
\begin{align}
S\to S+\bar{t}^{\bar{i}}\oint G^+\oint \bar{G}^+\bar{\phi}_i,\label{perturbact}
\end{align}
should be trivial, since the action is altered by a BRST-exact operator. In particular, an anti-holomorphic derivative with respect to the parameter $\bar{t}^{\bar{i}}$ only leads to the insertion of the operator $\oint G^+\oint \bar{G}^+\bar{\phi}_i$ in the topological partition function and should simply yield a vanishing result. However, this is not true because the so-called \textbf{holomorphic anomaly} spoils this reasoning \cite{Bershadsky:1993cx,Bershadsky:1993ta}: Inspecting the expression 
\begin{align}
\partial_{\bar{i}}F_g=\int_{\mathcal{M}_g}\langle |\prod_{a=1}^{3g-3}G^-(\mu_a)|^2\oint G^+\oint \bar{G}^+\bar{\phi}_i\rangle,
\end{align}
one can contour-deform the surface integrals $\oint G^+$ and $\oint \bar{G}^+$, which when hitting one of the $G^-$ and ${\bar G}^-$ give an insertion of the energy-momentum tensor
 \begin{align}
\partial_{\bar{i}}F_g=\int_{\mathcal{M}_g}\langle |\prod_{a=1}^{3g-4}G^-(\mu_{a})T(\mu_{3g-3})|^2\bar{\phi}_i\rangle,
\end{align}
up to an irrelevant numerical coefficient which we dropped.

\begin{figure*}[htbp]
\begin{center}
\epsfig{file=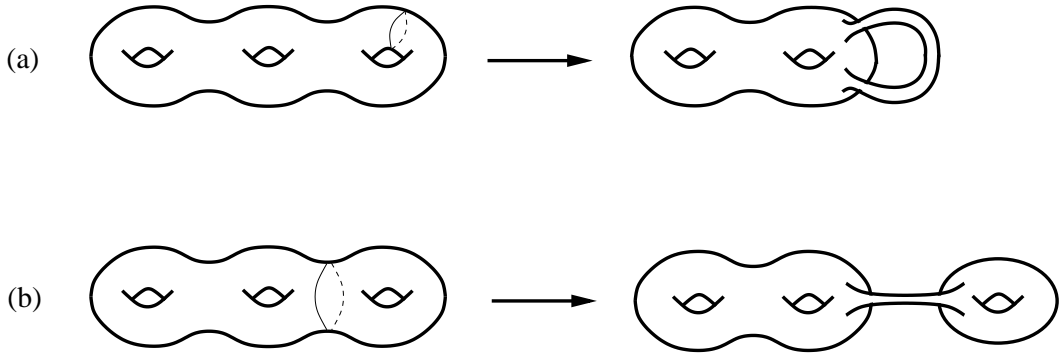, width=14cm}\vskip -0.3cm
\caption{Boundaries of a Riemann surface in moduli space: (a) pinching of a handle, (b) pinching of a dividing geodesic. In the limit of infinite length of the connecting tubes they are replaced by punctures at their end-points. A non-vanishing result is only achieved, if the states propagating on the tubes cancel the background charge for the sphere. In this case, they simply resemble the Yukawa couplings $C_{ijk}$.}\vskip -0.5cm
\label{fig:modbound}
\end{center}
\end{figure*}

The energy-momentum tensor insertion can be re-written as a partial derivative with respect to the world-sheet moduli, leading to a boundary contribution
 \begin{align}
\partial_{\bar{i}}F_g=\int_{\mathcal{M}_g}\frac{4\partial^2}{\partial m_b\partial\bar{m}_b}\langle |\prod_{a\neq b}G^-(\mu_{a})|^2\bar{\phi}_i\rangle.
\end{align}
A boundary in the moduli space of a compact Riemann surface corresponds to a surface with nodes, which appear either by pinching a handle or a dividing geodesic, as shown in Figure \ref{fig:modbound}.
In both cases, the surface develops a long and thin tube, which eventually in the pinching limit is turned into two punctures. Computing explicitly the contribution stemming from these two surfaces, the result is given by
\begin{align}
\partial_{\bar{i}}F_g=&\frac{1}{2}C_{\bar{i}\bar{j}\bar{k}}e^{2K}G^{j\bar{j}}G^{k\bar{k}}\bigg(\sum_{h=1}^{g-1}D_j F_hD_k F_{g-h}+\nonumber\\
&+D_jD_kF_{g-1}\bigg),\label{holoanom}
\end{align} 
where $D_i$ is the K\"ahler covariant derivative with respect to the chiral modulus field $\phi_i$, and $C_{\bar{i}\bar{j}\bar{k}}$ are the Yukawa couplings given by the three-point function of three chiral primaries on the sphere
\begin{align}
C_{ijk}=\langle\phi_i\phi_j\phi_k\rangle.
\end{align}
$K$ is the K\"ahler potential and $G_{i\bar j}=\partial_i\partial_{\bar j}K$ is the moduli metric related to $C$'s by special geometry.
An important point is that although the right hand side of (\ref{holoanom}) is not zero, it only contains contributions of lower genus $F_g$'s. Hence, the holomorphic anomaly leads to a recursion relation, which can be shown to be strong enough to yield the non-holomorphic part for all $F_g$'s iteratively \cite{Bershadsky:1993cx}.

\section{Topological Amplitudes and $\mathcal{N}=2$ Higher Derivative $\mathcal{F}$-terms}
{\it Certain  correlation functions of the gravitational sector of type II string-theory compactified on Calabi-Yau threefolds are shown to be computed by the partition function of the $\mathcal{N}=2$ topological string.}
\subsection{Physical Couplings and Topological Amplitudes}
After having introduced the topological partition function $F_g$, a further task is to make contact with the $\mathcal{N}=2$ compactifications of type II string theory.\footnote{We consider the case of $\mathcal{N}=(1,1)$ supersymmetry.} We are hence trying to find string scattering amplitudes (possibly at $g$-loop level), which can be reduced to $F_g$. Before plunging into the explicit calculation, let us first give a flavor of how this can happen.

The computation of scattering amplitudes in string-theory\footnote{Throughout these lectures we will use the RNS (Ramond-Neveu-Schwarz) formalism to compute amplitudes.} generically contains contributions from the space-time part of the vertex operators as well as from the internal (compact) sector and the (super)ghosts. After performing a number of computational steps (including spin-structure sum), a generic scattering amplitude is roughly of the factorized form \cite{N2}
\begin{align}
A\simeq &\frac{\langle(\text{space-time operators}) \rangle}{(\text{det}(\text{Im}\tau))^{d/2}}\cdot\nonumber\\
&\cdot\langle (\sigma\text{-model of the internal theory})\rangle,\label{musterequ}
\end{align} 
where $d$ is the number of non-compact space-time dimensions and $\tau$ is the period matrix of the (closed string) world-sheet. 
Although in general all these quantities are present, in the cases we are after, the space-time part cancels completely the $(\text{det}(\text{Im}\tau))$-factors and the remaining expression is only determined by the zero-mode part of the internal $\sigma$-model. Thus, it can be identified with some correlation function of the topological string (in the $\mathcal{N}=2$ case with the partition function).
\subsection{$\mathcal{N}=2$ Higher Derivative $\mathcal{F}$-terms}
Let us now study precisely for which type of string amplitudes the above mentioned cancellation occurs.
We consider type II theory compactified on a Calabi-Yau threefold. Thus, the underlying $\mathcal{N}=2$ SCFT has central charge $\hat{c}=3$ entailing that in order to obtain a non-vanishing result our scattering amplitudes have to balance a background charge of $3(g-1)$.

Approaching the problem in a systematic way \cite{N2}, we consider which operators are necessary to balance this charge. In the absence of external vertices, there are $(2g-2)$ insertions of picture changing operators (PCO) which stem from the integration over supermoduli of the world-sheet. Since they will only contribute with their internal part (in this case $G^-\bar{G}^-$), they make up already $(2g-2)$ units of the background charge. The remaining $(g-1)$ have to come from additional PCOs balancing the super-ghost charge of some additional vertex operators\footnote{For definition of vertex operators see Appendix \ref{app:vertex}.}. Since we want to compute amplitudes of the gravitational sector, one possibility to provide a total super-ghost charge of $-(g-1)$ is to insert $(2g-2)$ graviphotons in the $\left(-1/2\right)$-ghost picture. In order to cancel also the charge associated to space-time fermionic coordinates, we choose $(g-1)$ of them to come with a positive 4d helicity (+ sign in the 1st and 2nd plane) and the other half to come with a negative helicity. The correlation function is then schematically of the form
\begin{align}
\langle &\left(V^{\left(-1/2\right)}(T_{--})\right)^{g-1}\left(V^{\left(-1/2\right)}(T_{++})\right)^{g-1}\cdot\nonumber\\
&\cdot V_{\text{PCO}}^{3g-3}\rangle.
\end{align}
However, it turns out that this amplitude gives a vanishing result because of supersymmetry\footnote{Technically, this manifests itself in the fact that the sum over spin structures gives zero result.}. Thus, we insert two additional gravitons in the 0-ghost picture:
\begin{align}
\langle &\left(V^{(0)}(G)\right)^2\left(V^{\left(-1/2\right)}(T_{--})\right)^{g-1}\cdot\nonumber\\
&\cdot \left(V^{\left(-1/2\right)}(T_{++})\right)^{g-1}V_{\text{PCO}}^{3g-3}\rangle.\label{schemamplitude}
\end{align}

Before we explicitly evaluate this expression, let us examine more closely what term of the effective action this amplitude actually represents. The gravitons as well as the graviphotons are part of the 4 dimensional $\mathcal{N}=2$ supergravity multiplet. Supplemented by two spin $3/2$ gravitini, the component expansion of this Weyl chiral superfield reads \cite{deRoo:1980mm,Bergshoeff:1980is}
\begin{align}
{W_{a}}^{b,ij}=&{T_{a}}^{b,ij}+\theta^{c,[i}{\psi^{j]}_{ca}}^b+\nonumber\\
&+{(\sigma^{\mu\nu})_a}^b\theta^{i,c}{(\sigma^{\rho\tau})_c}^d\theta^j_dR_{\mu\nu\rho\tau},
\end{align}
where $a,b,c,d=1,2$ are (chiral) spinor indices, $i,j=1,2$ are $SU(2)$ indices labeling the supersymmetries and $\mu,\nu,\rho,\tau=0,\ldots,3$ are 4 dimensional space-time indices. On can then construct the following $1/2$-BPS $\mathcal{F}$-term in the effective action:
\begin{align}
\int d^4\theta ({W_{a_1}}^{b_1,ij}{W_{a_2}}^{b_2,kl}\epsilon^{a_1a_2}\epsilon_{b_1b_2}\epsilon_{ij}\epsilon_{kl})^g.
\label{W2gII}
\end{align}
Upon performing the superspace integration, this action term yields (among other contributions) precisely one term which corresponds to the amplitude (\ref{schemamplitude}), involving two self-dual Riemann tensors and $(2g-2)$ self-dual graviphoton field strengths, $R_+^2T_+^{2g-2}$.

We can now return to the explicit evaluation of the amplitude (\ref{schemamplitude}). This is done using the RNS formalism \cite{Green:1987sp} and performing the spin structure sum. The result is \cite{N2}:
\begin{align}
F_g=&\frac{\prod_{i=1}^{g}\int d^2x_i\prod_{j=1}^{g}\int d^2y_j}{(\text{det}(\text{Im}\tau))^2}|\text{det}\omega_i(x_j)|^2\cdot\nonumber\\
&\cdot|\text{det}\omega_i(y_j)|^2\int_{\mathcal{M}_g}\langle |\prod_{a=1}^{3g-3}G^-(\mu_{a_1})|^2\rangle,\label{resulta}
\end{align} 
where $\mu_a$ are the $(3g-3)$ Beltrami differentials parameterizing changes in the moduli of a genus $g$ Riemann surface. Comparing to (\ref{musterequ}), the space-time correlation function is:
\begin{align}
\prod_{i=1}^{g}\int d^2x_i\prod_{j=1}^{g}\int d^2y_j|\text{det}\omega_i(x_j)|^2|\text{det}\omega_i(y_j)|^2,
\end{align}
where $\omega$ are the $g$ abelian differentials defined on the world-sheet. The points $x_i$ and $y_i$ are the insertion points of the self-dual graviphoton and graviton vertex-operators of the two respective helicities, and the integration is understood over the homology cycles. It can be shown, that these integrals give (up to a constant factor which we neglect)
\begin{align}
\prod_{i=1}^{g}&\int d^2x_i\prod_{j=1}^{g}\int d^2y_j|\text{det}\omega_i(x_j)\text{det}\omega_i(y_j)|^2\simeq\nonumber\\
&\simeq (\text{det}(\text{Im}\tau))^2,
\end{align}
canceling as advertised the corresponding space-time zero-mode contribution in the denominator of (\ref{resulta}). The final expression is hence given by the topological expression
\begin{align}
F_g=\int_{\mathcal{M}_g}\langle |\prod_{a=1}^{3g-3}G^-(\mu_{a})|^2\rangle.\label{resultb}
\end{align} 
As already mentioned, the remarkable fact about this result is that it is precisely the partition function of the $\mathcal{N}=2$ topological string defined in (\ref{toppartition}).
\subsection{Non-renormalization Theorem}\label{sec:nonren}
In order to study possible perturbative or non-perturbative corrections to the above-mentioned topological amplitudes, it is necessary to figure out the precise dilaton dependence of the action term
\begin{align}
\sqrt{G}R^2T^{2g-2}.\label{topactterm}
\end{align}
In the string frame, this expression is proportional to
\begin{align}
\underbrace{e^{2\varphi(g-1)}}_{\text{loop-factor}}\cdot\underbrace{e^{2\varphi(g-1)}}_{\text{RR-insertions}}=e^{2\varphi(2g-2)}.\label{dilaton1}
\end{align}
The second factor stems from the fact that the graviphotons come from the Ramond-Ramond sector and are therefore accompanied by factors of the string coupling $g_s=e^{\varphi}$, with $\varphi$ the 4d dilaton field. Switching to the Einstein frame, the metric becomes dependent on the string coupling. The precise relation is derived by studying the Einstein-Hilbert term
\begin{center}
\begin{tabular}{ccc}
string frame & \large $\to$\normalsize & Einstein frame\\
$\sqrt{G}Re^{-2\varphi}$& \large $\to$\normalsize & $\sqrt{G}R$
\end{tabular}
\end{center}
Therefore, we read off that the metric is rescaled as
\begin{align}
G\to G e^{2\varphi}.
\end{align}
To establish the full dilaton dependence, we now have to count the metrics contained in (\ref{topactterm}). The measure factor $\sqrt{G}$ behaves like $G^2$ just like the Riemann tensor. To contract two Riemann tensors, one needs four inverse metrics, which yields another factor of $(G^{-4})$. Finally, to contract the graviphotons, another $(2g-2)$ inverse metrics are deployed giving $G^{-(2g-2)}$. All together, they behave as $G^{-(2g-2)}$, yielding an additional contribution of $e^{2\varphi(-2g+2)}$ which precisely cancels (\ref{dilaton1}). Hence, we see that there is no dilaton dependence in the Einstein frame at all.

This result is consistent with the well known fact that in $\mathcal{N}=2$ type II compactifications in 4 dimensions, the dilaton resides in a hypermultiplet which does not couple to vector multiplets through local effective action terms. From this, we conclude that there should be no corrections (perturbative or not) to the $F_g$'s given by the expression (\ref{resultb}).
\subsection{Duality Mapping}
The starting point for a duality mapping between heterotic string and type IIA is in 6 dimensions. There, it was observed that the moduli space of the heterotic string compactified on $T^4$ is identical to that of type IIA compactified on $K3$, and the two theories can be identified upon inversion of the six-dimensional (6d) string coupling. From this result one can derive further dualities by compactifying down to 4 dimensions. In particular, one can establish a duality between type IIA on a $K3$-fibered Calabi-Yau threefold and the heterotic string on $K3\times T^2$. Keeping track of the 4d dilaton, which resides in an $\mathcal{N}=2$ hypermultiplet on the type II side, one finds that it belongs to an $\mathcal{N}=2$ vector multiplet on the heterotic side. The reason is that the heterotic dilaton is mapped to the volume of the base of the $K3$-fibration in the type II side. For the $F_g$'s, this means that they are exact on the IIA side (as mentioned above) and appear already at genus $g=1$ on the heterotic side, where however they receive corrections. Despite its non exactness, their one-loop heterotic representation is very useful and allows many properties to be uncovered and studied in a simple way~\cite{AGNThet}. 
\section{$\mathcal{N}=1$ Topological Amplitudes}
{\it A reduction of the $\mathcal{N}=2$ topological amplitudes to $\mathcal{N}=1$ is achieved in two ways: (i) the heterotic dual is studied, being captured by a torus integral; (ii) $\mathbb{Z}_2$ world-sheet involutions of the closed string amplitudes are found to result in type I open string amplitudes. For both theories the analog of the holomorphic anomaly equation is derived.}
\subsection{Heterotic String Theory}
\subsubsection{Topological Amplitudes}\label{topamp}
We choose the left moving side of the heterotic string to be the supersymmetric one and thus, the topological twist will only be performed on this side. To match the amplitudes of the type IIA side, we have to insert $(2g-2)$ left-moving spin fields. Possible candidates are gauginos, gravitinos or dilatinos. For reasons which will become clear below, we will opt for the first ones and we will supplement them by two gauge fields. Thus, the transition from the type IIA side is given by 
\begin{align}
\text{graviphoton }&\to \ \text{gaugino},\nonumber\\
\text{graviton }&\to \ \text{gauge field},\label{conversion}\\
\mathcal{N}=2 \ \text{Weyl }{W_a}^{b,ij}\ &\to \ \mathcal{N}=1\ \text{gauge }W_\alpha^a,\nonumber
\end{align}
which is given by the chiral multiplet
\begin{align}
W^\alpha_a=-i\lambda_a^\alpha-\frac{i}{2}{(\sigma^\mu\bar{\sigma}^\nu)_a}^b\theta_b F^\alpha_{\mu\nu}+\ldots.\label{gaugemulti}
\end{align}
As before $a,b=1,2$ are spinor indices, while $\alpha$ labels the gauge group factor.

The corresponding expression to the type IIA effective action term (\ref{W2gII}) is the following $1/2$-BPS $F$-term~{\cite{Antoniadis:1996qg}}:
\begin{align}
\int d^2\theta F_g^{\text{HET}}(\text{Tr}W^2)^g=F_g^{\text{HET}}F^2(\lambda\lambda)^{g-1}.
\end{align}
An additional complication arising here is that besides the $\text{det}(\text{Im}\tau)$ factors, there are extra contributions stemming from double poles of the operator product expansion of the right-moving gauge charge operators involved in the amplitude. In order to get rid of those contact terms, a suitable difference of two gauge groups with no charge matter must be taken:
\begin{align}
\int d^2\theta F_g^{\text{HET}}[(\text{Tr}_1-\text{Tr}_2)W^2]^g,
\end{align}
where $\text{Tr}_i$ denotes tracing with respect to the $i$-th gauge group factor. Thus, finally, only the Kac-Moody currents of the vertex operators are left to deal with. These contribute only with their zero-mode part, given by 
\begin{align}
\sum_{i=1}^gQ_i^a\bar{\omega}_i,
\end{align}
where $Q_i^a$ is an operator measuring the $a$-th charge of the state that propagates on the $i$-th loop of the world-sheet, and the abelian differentials $\bar\omega$ provide the missing anti-holomorphic contributions to cancel (upon integration) the $\text{det}(\text{Im}\tau)$-factors.

With these subtleties, the topological amplitudes take the form \cite{Antoniadis:1996qg}
\begin{align}
F_g^{\text{HET}}=\int _{\mathcal{M}_g}&\langle\prod_{a=1}^{3g-3}(\mu_a G^-)(\bar{\mu}_a\bar{b})(\text{det}Q_i^{b_j})\cdot\nonumber\\
&\cdot(\text{det}Q_i^{c_j})\rangle.
\label{Fghet}
\end{align}
\subsubsection{Holomorphic Anomaly Equation}
An interesting puzzle appears when trying to repeat the computation leading to the holomorphic anomaly equation. After perturbing the action by the BRST exact operator\footnote{Note, that only one BRST operator is present in contrast to two of them in (\ref{perturbact}).}
\begin{align}
\{Q_{\text{BRST}},\bar{\phi}_{\bar{i}}\},
\end{align}
a derivative with respect to an anti-holomorphic modulus simply leads to an insertion of this operator in the amplitude. The naive vanishing of the resulting expression is again spoiled by the appearance of (anomalous) boundary terms, which take the form
\begin{align}
\partial_{\bar{i}}F^g=\sum_{g_1+g_2=g}F^{g_1}_{\bar{i},\bar{j}}G^{\bar{j}j}D_jF^{g_2}.
\end{align}
The new feature of this equation, compared to the corresponding one in type IIA theory, is the appearance of additional objects on the right hand side which are of the form \cite{Antoniadis:1996qg}
\begin{align}
F^g_{\bar{i},\bar{j}}=\int_{\mathcal{M}_{g,n}}&\langle\prod_{a=1}^{3g-3+n}(\mu_a G^-)(\bar{\mu}_a\bar{b})(\text{det}Q_i^{b_j})\cdot\nonumber\\
&\cdot(\text{det}Q_i^{c_j})\int \bar{\phi}_{\bar{i}}\int \tilde{\bar{\phi}}_{\bar{j}}\rangle.
\end{align}
These new objects indicate that the recursion relations do not close within the topological partition function, but one has to allow for operators in the twisted theory, which are not in the kernel of the (twisted) BRST operator $Q_{\text{BRST}}$.  

The additional terms can be computed from
\begin{align}
F^g_{n}\Pi^nW^{2g},
\end{align}
where the $F^g_{n}$ are arbitrary functions of chiral superfields and the $\Pi$'s denote the chiral projection of non-holomorphic functions of chiral superfields.\footnote{In this sense it can be viewed as a generalization of the $\bar{D}^2$ operator of rigid supersymmetry.} The component expansion of $\Pi$ is given as
\begin{align}
\Pi=\bar{\lambda}\bar{\lambda}+\theta^2((\partial \bar{Z})^2+\partial^2\bar{Z})+\ldots ,
\label{Piexpansion}
\end{align}
where $Z$ and $\lambda$ denote the bosonic and fermionic components of chiral superfields. However, the introduction of these new objects alone does not seem to solve the problem of integrability, which remains open.
\subsection{Type I}\label{typeI}
\subsubsection{$\mathbb{Z}_2$ Involution}
The simplest way to compute open string scattering amplitudes is to take $\mathbb{Z}_2$ involutions of closed (type IIB) string amplitudes \cite{Angelantonj:2002ct,Blau:1987pn,Bianchi:1988fr}. This is done by identifying homology cycles of the world-sheet and taking an appropriate ``square root''. In this manner, the previously closed world-sheet gets a number of boundaries, which are precisely the fixed points of the involution, and the $F_g$'s are generalized to $F_{(g,h)}$ (in the simplest case of oriented surfaces). We will preliminary focus on world-sheets which have no remaining handles. Let us study this case more carefully with the following example:\\[10pt]
\textbf{Example}
{\it We consider a genus 3 Riemann surface as the double cover (that is as `the square') of a Riemann surface with no handles but 4 boundaries (see Figure \ref{fig:genus3}).
\begin{figure}[htbp]
\begin{center}
\epsfig{file=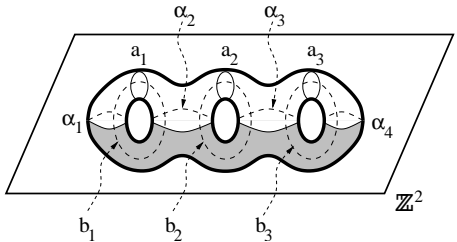, width=6cm}\vskip -0.5cm
\caption{Example of a $\mathbb{Z}_2$ involution of a genus 3 Riemann surface. The $(\mathbf{a_i},\mathbf{b_i})$ are the canonical homology basis of the Riemann surface and the $\mathbf{\alpha_j}$ with $j=1,\dots,4$ form the boundaries of the open Riemann surface. Note, that upon different identifications of the homology cycles different open surfaces appear.}\vskip -1cm
\label{fig:genus3}
\end{center}
\end{figure}
The canonical homology basis for this surface is given by the cycles $(\mathbf{a_i},\mathbf{b_i})$, with $i=1,2,3$, while the boundaries of the open string surface obtained upon a special $\mathbb{Z}_2$ involution are the cycles $\alpha_1,\ldots,\alpha_4$. From Figure \ref{fig:genus3} we read off that they are given by
\begin{eqnarray}
\mathbf{\alpha_1}&=&\mathbf{a_1},\nonumber\\ \mathbf{\alpha_2}&=&\mathbf{a_2a_1^{-1}},\nonumber\\ \mathbf{\alpha_3}&=&\mathbf{a_3a_2^{-1}},\label{cycles}\\ \mathbf{\alpha_4}&=&\mathbf{a_3^{-1}}.\nonumber
\end{eqnarray}
The special $\mathbb{Z}_2$ involution which is displayed in Figure \ref{fig:genus3} is defined to act on the cycles as follows
\begin{align}
&\mathbf{a_i}\to \mathbf{a_i},&& \mathbf{b_i}\to -\mathbf{b_i}.\label{examptrans}
\end{align}}
\subsubsection{Scattering Amplitudes}
We now consider a surface with $h$ boundaries, that is obtained from the $\mathbb{Z}_2$ involution of a closed Riemann surface of genus $g=h-1$. For simplicity, we will focus on the case where all vertex operator insertions are on the boundaries \cite{Antoniadis:2005sd}. In this case, in order to make contact with the type II calculation, we have to use two gauge fields and $2h-4$ gauginos. The reason is that upon the $\mathbb{Z}_2$ involution the graviton becomes a gauge field and the graviphoton a gaugino, as described in Section \ref{topamp}. One then finds the physical coupling
\begin{align}
\int d^2\theta F_{(0,h)}(\text{Tr}W^2)^{h-2},
\end{align}
where $W$ is the gauge supermultiplet defined in (\ref{gaugemulti}).
Since the vertex insertions are path ordered, the insertion of more than two on the same boundary would lead to rather complicated expressions. Hence, we distribute the fields as follows
\begin{itemize}
\item on each of the $h-2$ boundaries we insert two gauginos
\item on one boundary we insert the two gauge fields
\item one boundary remains `empty' and serves as a spectator
\end{itemize}
Using furthermore Neumann boundary conditions\footnote{This choice is dictated by the necessity of canceling the $\text{det}(\text{Im}\tau)$ factors.} the amplitudes are given by
\begin{align}
F_{(0,h)}=\int_{\mathcal{M}_h}\langle\prod_{a=1}^{3h-6}\int (\mu_aG^-)\rangle\cdot(\text{lattice sum}).
\end{align}
It can be shown that this result is dual to (\ref{Fghet}) by heterotic/type I duality.
\subsubsection{Holomorphic Anomaly Equation}
\begin{figure*}[htbp]
\begin{center}
\epsfig{file=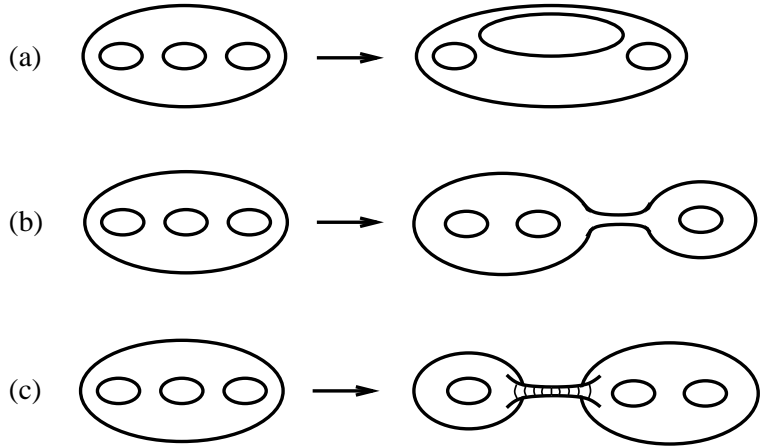, width=10cm}
\caption{Boundaries of an open string world-sheet.}
\label{fig:openboundaries}
\end{center}
\end{figure*}

Performing the same computation as before and taking anti-holomorphic derivatives of the amplitude, one has to consider again boundary contributions in the moduli space of open string world-sheets \cite{Antoniadis:2005sd}. In contrast to the type II computation however, this time one has to consider three cases, which are displayed in Figure \ref{fig:openboundaries}. While in the first two cases the surface develops a long and thin strip, in the last case it develops a cylinder which corresponds to a closed string exchange between the two new surfaces.
Just as in the heterotic case, there appear new objects when taking an anti-holomorphic derivative of $F_{(0,h)}$, spoiling the closure of the recursion relation. 
\subsubsection{Outlook to $F_{(g,h)}$}
A question which immediately arises is whether it is possible to relax the constraint of $g=0$. In other words, are there physical string couplings which correspond to $F_{(g,h)}$ with $g\neq 0\neq h$? The main problem in this respect is that due to the lower number of boundaries, a distribution of the vertex insertions as in $F_{(0,h)}$ is not possible. One has to either put more than two vertices on some boundaries, or consider field insertions in the bulk. In both options the position integrals do not quite cancel the $(\text{det}(\text{Im})\tau)$ factors which arise from the space-time part. Thus, these amplitudes are not precisely topological.
\section{Genus 2 Amplitudes, Supersymmetry Breaking and Fermion Masses}
{\it The $\mathbb{Z}_2$ involutions of genus 2 topological amplitudes describe the communication of supersymmetry breaking in some ``hidden" sector to a gauge theory living on the boundary. More precisely, they compute the resulting gaugino or fermion masses.}
\subsection{Overview}\label{overview}
\begin{figure*}[htbp]
\begin{center}
\epsfig{file=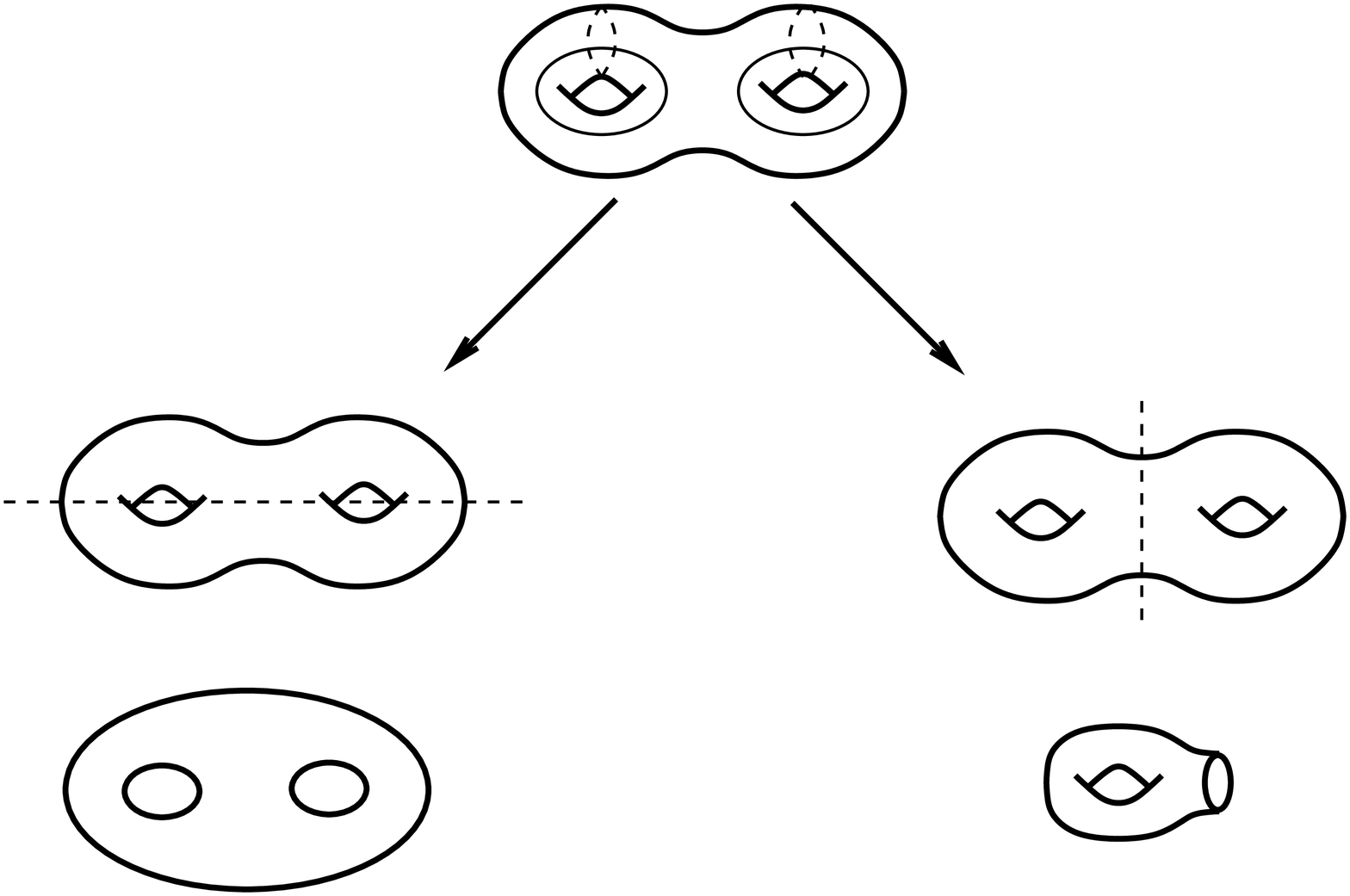, width=11cm}\\ [8pt]
$Z_1=\left(\begin{array}{cccc} 1 & 0 & 0 & 0 \\ 0 & 1 & 0 & 0 \\ 0 & 0 & -1 & 0 \\ 0 & 0 & 0 & -1 \end{array}\right)\hspace{3.5cm}$  $Z_2=\left(\begin{array}{cccc} 0 & 1 & 0 & 0 \\ 1 & 0 & 0 & 0 \\ 0 & 0 & 0 & -1 \\ 0 & 0 & -1 & 0 \end{array}\right)$\vskip -0.5cm
\caption{Possible involutions of a genus 2 compact Riemann surface and their involution matrices acting on the homology cycles $C=(\mathbf{a_1},\mathbf{a_2},\mathbf{b_1},\mathbf{b_2})$. For instance $Z_1C^T$ reproduces the transformation (\ref{examptrans}).}\vskip -0.5cm
\label{fig:genus2}
\end{center}
\end{figure*}

Studying in detail the genus 2 case, there are two possible inequivalent involutions, which together with their involution matrices are schematically displayed in Figure \ref{fig:genus2}. 
These involutions lead to a different identification of cycles of the homology basis, corresponding to a different form of the involution matrix. Furthermore, the field insertions on the closed world-sheet are initially two gravitons ($R$) and two graviphotons ($T$). Depending on whether they are inserted in the bulk or on the boundary, they either stay gravitons and graviphotons or get converted to gauge fields ($F$) and gauginos ($\lambda$), respectively, according to (\ref{conversion}). Both diagrams communicate a breaking of supersymmetry in some hidden sector of the theory (e.g. in the bulk or in some boundary) to the ``observable" sector (e.g. a gauge theory living on some supersymmetric --to lowest order-- boundary).

Indeed, the diagram to the left of Figure~\ref{fig:genus2} has three boundaries with two gauginos on one of them and two gauge fields on another, giving rise to $F_{(0,3)}$. Replacing the gauge field strengths by their $D$-term auxiliary components, one obtains a gaugino mass when supersymmetry is broken by a $D$-term expectation value. This provides an example of gauge mediation~\cite{Antoniadis:2005sd} that will be studied below, in subsection \ref{sec:gaugemed}. It is also worth noticing that the holomorphic anomaly of $F_{(0,3)}$ brings a term $\Pi W^2$, generated at one loop from the annulus diagram. Using the component expansion (\ref{Piexpansion}), and replacing the gauge field strengths by $D$-term auxiliary components, one obtains fermion masses of the type $\mu$-term, needed in the supersymmetric standard model.

On the other hand, the diagram to the right of Figure \ref{fig:genus2} has one handle and one boundary with two gauginos. Moreover, there is a graviton insertion in the bulk that can be replaced by its auxiliary component, generating again a gravitino mass related to $F_{(1,1)}$, upon supersymmetry breaking in the gravity sector~\cite{Antoniadis:2004qn}. Thus, this provides an example of gravity mediation that we study next, in the case where supersymmetry breaking arises by compactification via the Scherk-Schwarz (SS) mechanism \cite{Scherk:1978ta}.
\subsection{Scherk-Schwarz (SS) Mechanism for Supersymmetry Breaking}
A possibility to break supersymmetry spontaneously with the help of compact dimensions in field theory was first introduced in \cite{Scherk:1978ta}. Considering (in the simplest case) compactification on a circle of radius $R$, one can allow the fields to be periodic only up to an $R$-symmetry transformation
\begin{align}
&\Phi(y+2\pi R)=U\Phi(y), & \text{with} && U=e^{2\pi iQ}\, ,
\label{period}
\end{align}
where $Q$ is the corresponding $R$-charge. Since, upon compactification, $R$-symmetry is restricted to discrete subgroups (that is $U^N=1$ for $\mathbb{Z}_N$), $Q$ is quantized. Moreover, from the periodicity (\ref{period}), one finds that the Kaluza-Klein (KK) momentum takes the form
\begin{align}
p=\frac{n+Q}{R},\label{SSstates}
\end{align}
for integer $n$, which leads to a mass-shift of the KK modes by $Q/R$. In particular, the gravitino zero mode gets a mass $Q/R$ and (local) supersymmetry is broken.

This method is generalized in closed string theory using world-sheet modular invariance~\cite{Rohm:1983aq,Ferrara:1988jx}. For open strings, the effect is identical to field theory, for Neumann boundary conditions. For Dirichlet conditions on the other hand, describing a D-brane perpendicular to the SS coordinate, supersymmetry remains unbroken, since there are no KK modes~\cite{Antoniadis:1998ki}.
\subsection{Gravity 
Mediated Supersymmetry Breaking}
\begin{figure}[htbp]
$~$\hskip 1.3cm\epsfig{file=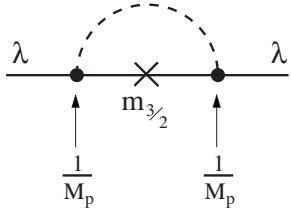, width=7cm}\vskip -0.6cm
\caption{Field theory 2-point function of two gauginos at zero-momentum.} 
\vskip -0.8cm
\label{fig:fieldth}
\end{figure}

Consider a generic 1-loop two-point function with some gravitational exchange in the effective field theory (see Figure \ref{fig:fieldth}).
Since each of the two gravitational vertices comes with a factor of the inverse Planck mass, $M_p^{-1}$, the generic form of the gaugino mass $m_{1/2}$ is 
\begin{align}
m_{1/2}\sim \frac{m_{3/2}}{M_p^2}\times \left\{\begin{array}{ll} \Lambda^2_{\text{UV}}
\\ &\\ m_{3/2}^2
\end{array}\right.
\label{m12FT}
\end{align}
where the gravitino mass insertion $m_{3/2}$ is obviously needed for a non-zero result. The form of the extra factor in the right hand side is determined by dimensional analysis and depends on whether the loop integral is (quadratically) divergent or convergent. In the former case, identifying the ultraviolet (UV) cutoff $\Lambda_{\text{UV}}$ with $M_p$, one obtains $m_{1/2}\sim m_{3/2}$. In the second case, $m_{1/2}$ is hierarchically smaller than the gravitino mass in the limit $m_{3/2}<<M_p$.

\begin{figure}[htbp]
\begin{center}
\epsfig{file=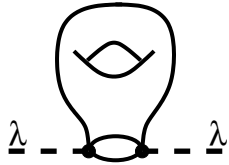, width=3cm}\vskip -0.5cm
\caption{Two-point function of two gauginos at 1-loop level leading to a mass-term.}\vskip -0.5cm
\label{fig:2gauginos}
\end{center}
\end{figure}

Now one can study the two-point function of Figure \ref{fig:2gauginos} for type II compactifications on $T^2\times K3$, where the SS circle of radius $R$ is embedded in $T^2$. 
The simplest approach for the computation would be to realize $K3$ as some torus orbifold $T^4/\mathbb{Z}_N$, however, then the gaugino mass is protected by the orbifold symmetries and the loop integral vanishes. This technical problem can be avoided by blowing up the orbifold singularity of $K3$. The result is then found to be \cite{Antoniadis:2004qn}
\begin{align}
m_{1/2}\propto g_s^2\frac{m_{3/2}^3}{M_s^2},
\label{gravimed}
\end{align} 
in the limit $m_{3/2}<<M_s$, where $M_s$ is the string scale, and the string coupling $g_s$ equals the square of the gauge coupling $\alpha_G$.
From this we conclude that the gravitational mediation of supersymmetry breaking, induced by the string diagram in Figure \ref{fig:2gauginos}, reproduces the field theory result (\ref{m12FT}) in the case of a UV convergent loop integral. The absence of the first term, with a linear behavior in $m_{3/2}$, implies also the absence of a contribution due to the so-called anomaly mediation. The latter would have exhibited a result of the form
\begin{align}
m_{1/2}\sim \alpha_G m_{3/2},
\end{align}
where, as we see, the powers of $g_s$ do not match with (\ref{gravimed}). This is in accordance with the fact that the one-loop result (notice the power of the gauge coupling constant) always vanishes. This is because the $\mathcal{N}=2$ superconformal $U(1)$ charge of two gaugino insertions cannot be balanced by one picture changing operator, as would be necessary from charge counting. Thus, one needs to consider a higher loop order, which however behaves as $m_{3/2}^3$, as seen above.

Focusing now on explicit numerical results, from (\ref{SSstates}), the gravitino mass is proportional to the inverse of the compactification radius
\begin{align}
m_{3/2}\sim1/R\, , 
\end{align}
associated to the SS interval perpendicular to the Standard Model branes. $R$ is then determined from the standard relation between the 4d Planck mass and the string scale:
where the value of the radius of the SS circle is given by
\begin{align}
R^{-1}=\frac{2M_s^3}{\alpha_G^2 M_p^2}\sim 10^{13}\text{GeV}.
\end{align}
From the result (\ref{gravimed}), one then finds that the gaugino mass is in the TeV range if every loop factor brings a typical suppression by two orders of magnitude. On the other hand, scalars on the brane acquire generically one-loop mass corrections from the annulus diagram~\cite{Antoniadis:1998ki}. It follows that
\begin{align}
m_0\geq g_s\frac{m_{3/2}^2}{M_s}\sim 10^8\text{GeV}\, .
\end{align}
Thus, this mechanism leads to a hierarchy between scalar and gaugino masses of the type required by split supersymmetry~\cite{split,splitstring}.
\subsection{Breaking Local Supersymmetry without Breaking $R$-symmetry}
\begin{figure*}[htbp]
\begin{center}
\epsfig{file=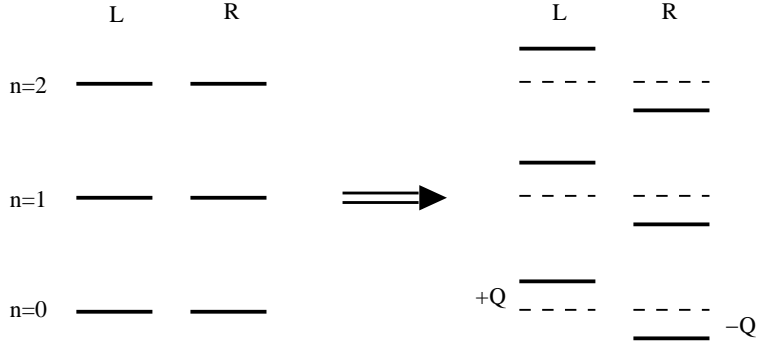, width=10cm}
\caption{Mass-shifts of the left- and right-handed spin-$3/2$ KK states due to SS supersymmetry breaking.}
\label{fig:SSshift}
\end{center}
\end{figure*}

We consider now the case where the contribution of gravity mediation of Figure~\ref{fig:2gauginos} is suppressed, such as in special models mentioned above with additional discrete symmetries. Another example is when the breaking of local supersymmetry preserves an $R$-symmetry protecting the vanishing of gaugino masses. This is provided again by the simple case of SS supersymmetry breaking using a $\mathbb{Z}_2$ $R$-parity, corresponding to $Q=1/2$ in (\ref{SSstates}). For generic $Q$, the masses of the spin-$3/2$ KK states get modified as displayed in Figure \ref{fig:SSshift}.
One sees that mass-shifts with a generic $Q$ lead to Majorana masses that break $R$-symmetry. However in the case of $Q=1/2$, one notices a pairing of the states
\begin{align}
&|n+Q\rangle_L, &&\text{and} &&|n+1-Q\rangle_R\, ,
\end{align}
forming Dirac spinors. Thus, in this case $R$-symmetry is left intact and gravity mediated corrections to gaugino masses vanish~\cite{splitstring}. One can then consider corrections arising from the diagram on the left of Figure~\ref{fig:genus2} with 3 boundaries, describing gauge mediation, that we study next.
\subsection{Gauge Mediated Supersymmetry Breaking}\label{sec:gaugemed}
\begin{figure*}[htbp]
\begin{center}
\epsfig{file=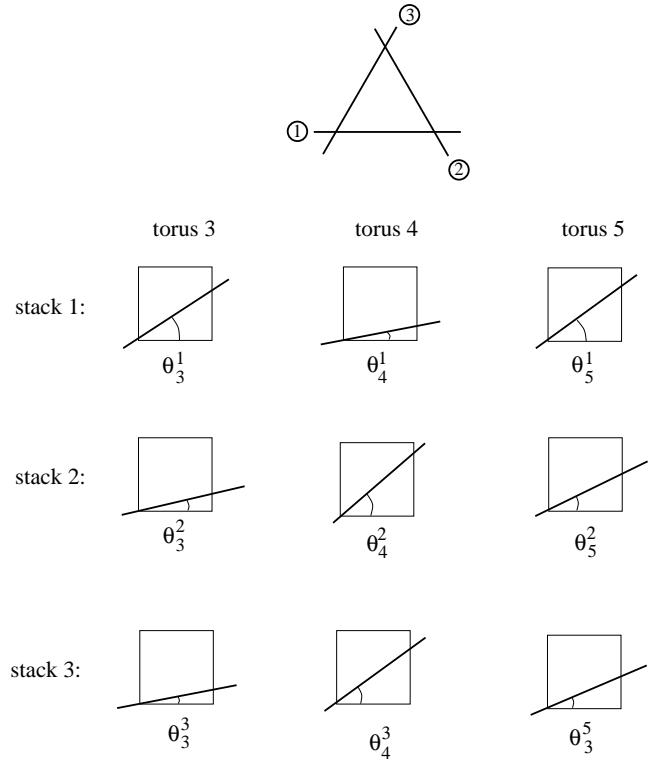, width=8.5cm}\vskip -0.5cm
\caption{Three stacks of branes at angles $\theta_I$ corresponding to the three tori. If all three branes would intersect at a single point in one torus, the diagram would vanish.}\vskip -0.5cm
\label{fig:branesatangles}
\end{center}
\end{figure*}

Realizing the Calabi-Yau threefold as an orbifold $\prod_{I=3}^5T^2_I$, the diagram with three boundaries 
corresponds to three stacks of magnetized D9-branes \cite{Antoniadis:2005sd}. The brane stacks are associated with the gauge groups $U(N_a)$ with $a=1,2,3$. Each of the three abelian factors contains an internal magnetic field $H^a_I$ on each $T^2_I$. An alternative description is obtained by T-dualizing one direction of each torus, resulting in a type IIA orientifold with three stacks of D6-branes at angles attached to the boundaries of the world-sheet, according to Figure~\ref{fig:branesatangles}.

The angles of the branes relative to the horizontal axis of each torus are given by
\begin{align}
\theta_I^a=\text{arctan}H^a_I\alpha',
\end{align}
and the condition for preserving $\mathcal{N}=1$ supersymmetry at each stack is that the sum of the angles over the three planes is zero
\begin{align}
\sum_{I=3}^5\theta_I^a=0,\hspace{1cm}\forall a=1,2,3.
\end{align}
The angles $\theta_I^a$ play the role of the orbifold twists around the $\mathbf{b}$-cycles, while the twists around $\mathbf{a}$-cycles are trivial (see Figures~\ref{fig:genus3} and \ref{fig:genus2}). For the special case at hand, we find
\begin{align}
&2\pi h_I^1=2(\theta_I^1-\theta_I^3), &&2\pi h_I^2=2(\theta_I^3-\theta_I^2),
\end{align}
with $h_I^i$ the orbifold twist in the $I$-th plane around the $i$-th $\mathbf{b}$-cycle. Here the boundaries in terms of the canonical homology basis are as in Figure \ref{fig:genus3}: 
$\mathbf{\alpha_1}=\mathbf{a_1}, 
\mathbf{\alpha_2}=\mathbf{a_2}^{-1}, 
\mathbf{\alpha_3}=\mathbf{a_1}^{-1}\mathbf{a_2}$.

The general non-supersymmetric case is obtained by breaking the relation for the angles adding up to zero by a small parameter $\epsilon$
\begin{align}
\sum_{I=3}^5\theta^a_i=\epsilon,
\end{align}
which is immediately translated into a relation for the orbifold twists: 
$\sum_{I=3}^5h^i_I=-2\epsilon$.
As described in Section~\ref{overview}, at the level of the effective action, the gaugino acquires a mass term via $D$-term supersymmetry breaking. This stems from the supersymmetric $\mathcal{F}$-term 
\begin{align}
(\text{Tr }W_1^2)(\text{Tr }W_2^2),\label{breakF}
\end{align}
by assigning an expectation value $m_0^2$ to the auxiliary field of the $U(1)$ vector multiplets\footnote{We consider the weak field limit of small $\epsilon$.}
\begin{align}
\langle D\rangle\sim m_0^2 =\epsilon M_s^2\, .
\end{align}
The two factors in (\ref{breakF}) correspond to the gauge groups $U(N_1)$ and $U(N_2)$, respectively. Taking the $D$-auxiliary components of one factor, one gets gaugino masses for the other gauge group factors, given by the topological partition function $F_{(0,3)}$. Indeed, the gaugino masses can be computed explicitly by the 2-point function at zero momentum.\footnote{For the definition of the vertex operators see Appendix \ref{app:vertex}.} In the limit $\epsilon\to 0$, the result is
\begin{align}
m_{1/2}=g_s^2F_{(0,3)}\frac{m_0^4}{M_s^3}\, ,
\end{align} 
in agreement with the field theory expectation. 

Note that one finds again a hierarchy between scalar and gaugino masses. In order to have the latter in the TeV scale, one has to introduce scalar masses of the order of $10^{13}-10^{14}$ GeV. On the other hand, as discussed in subsection~\ref{overview}, this mechanism provides also $\mu$-type fermion masses of the same order of magnitude as gauginos (up to a loop factor). Thus, one obtains again a string realization of split supersymmetry.
\section{$\mathcal{N}=4$ Topological Amplitudes}
{\it Compactifications on $K3$ and $K3\times T^2$ are considered and scattering amplitudes of the type II string-theory are found to be computed by correlation functions of the topological string. Since the partition function is vanishing due to the failure of canceling the background charge deficit, a regularization by suitable insertions of additional operators has to be considered. The focus will be on `minimal' couplings, mapped to as simple as possible loop calculations on the heterotic side. As an application of these amplitudes the harmonicity relation as a generalization of the $\mathcal{N}=2$ holomorphicity will be mentioned briefly.}
\subsection{Motivation}
After having studied topological amplitudes associated to $\mathcal{N}=2$ compactifications on Calabi-Yau threefolds, the question is whether one can obtain similar results for $\mathcal{N}=4$. The study of these amplitudes can be motivated by a number of reasons:
\begin{itemize}
\item $\mathbb{Z}_2$ involutions of $\mathcal{N}=4$ closed string amplitudes possibly lead to $\mathcal{N}=2$ amplitudes with boundaries, therefore providing a generalization of the presently known $F_g\equiv F_{(g,0)}^{\mathcal{N}=2}$. 
\item $\mathcal{N}=4$ couplings necessarily involve higher derivative corrections to the Einstein-Hilbert action of the type $R^4$, modifying possibly the entropy of \textbf{$\mathcal{N}=4$ black holes}.
\item The generalization of the $\mathcal{N}=2$ holomorphicity (up to a possible anomaly) to the \textbf{harmonicity relation} could provide interesting information on the structure of the effective $\mathcal{N}=4$ supergravity.
\end{itemize}
\subsection{Compactification on $K3$}
\subsubsection{$\mathcal{N}=4$ Superconformal Algebra}\label{N4SCFalg6D}
The critical dimension for compactifications on $K3$ is $\hat{c}=2$. Since this number is different from 3, it implies a trivial vanishing of the topological partition function. Indeed, the topological twist is performed using the $U(1)$ current of an $\mathcal{N}=2$ subalgebra of the full $\mathcal{N}=4$. The $U(1)$ charge of the $(3g-3)$ insertions of $G^-$ does then not match with the background charge deficit.

The idea, already pursued in \cite{BerVafa,conjectures}, is to consider instead a correlation function with an appropriate number of additional insertions of the second (complex) supercharge of the $\mathcal{N}=4$ SCFT algebra, $\tilde{G}^+$. The latter consists of two doublets of supercurrents ($(G_{K3}^+,\tilde{G}_{K3}^-)$ and $(\tilde{G}^+_{K3},G^-_{K3})$) and the generators of an $SU(2)$ affine current algebra ($(J_{K3}^{++},J_{K3},J^{--}_{K3})$), as well as their right moving counterparts. The algebra fulfilled by the left moving operators is determined by the short distance operator product expansion (OPE):
{\allowdisplaybreaks
\begin{eqnarray}
J^{--}_{K3}(z)J^{++}_{K3}(0)&\sim&\frac{J_{K3}(0)}{z},\nonumber\\
J^{--}_{K3}(z)G^+(0)&\sim&\frac{\tilde{G}^-_{K3}(0)}{z},\nonumber\\
J_{K3}^{++}(z)\tilde{G}_{K3}^-(0)&\sim&-\frac{G^+_{K3}(0)}{z},\nonumber\\
J_{K3}^{++}(z)G^-_{K3}(0)&\sim&\frac{\tilde{G}^+_{K3}(0)}{z},\nonumber\\
J_{K3}^{--}(z)\tilde{G}^+_{K3}(0)&\sim&-\frac{G^-_{K3}(0)}{z},\nonumber\\
G^+_{K3}(z)G^-_{K3}(0)&\sim&\frac{J_{K3}(0)}{z^2}+\nonumber\\
&&+\frac{2T_{K3}(0)+\partial J_{K3}(0)}{z},\nonumber\\
\tilde{G}^+_{K3}(z)\tilde{G}^-_{K3}(0)&\sim&\frac{J_{K3}(0)}{z^2}+\nonumber\\
&&+\frac{2T_{K3}(0)+\partial J_{K3}(0)}{z},\nonumber\\
G^+_{K3}(z)\tilde{G}^+_{K3}(0)&\sim&\frac{J^{++}_{K3}(0)}{z^2}+\frac{\partial J^{++}_{K3}(0)}{2z},\nonumber\\
G^-_{K3}(z)\tilde{G}^-_{K3}(0)&\sim&\frac{J^{--}_{K3}(0)}{z^2}+\frac{\partial J^{--}_{K3}(0)}{2z},\nonumber
\end{eqnarray}}
$~$\hskip -0.25cm with all other OPE vanishing and a similar set of relations for the right-movers.

For the topological twist, we use the $\mathcal{N}=2$ subsector $(T,G^{\pm},J)$ and proceed as in Section \ref{sec:toptwist}, by redefining the energy momentum tensor as in (\ref{toptwist}).
Just as in the $\mathcal{N}=2$ sector, the conformal weights of the operators are changed accordingly.
\begin{center}
\begin{tabular}{ccc}\hline
\textbf{operator} & \textbf{conf. weight} & \textbf{$U(1)$}\\ \hline
\parbox{0.5cm}{\vspace{0.1cm}$\tilde{G}^+$\vspace{0.1cm}} & \parbox{0.2cm}{\vspace{0.1cm}$1$\vspace{0.1cm}} & \parbox{0.2cm}{\vspace{0.1cm}$1$\vspace{0.1cm}} \\
$\tilde{G}^-$ & $2$ & $-1$\\
$J^{++}$ & $0$ & $2$\\
$J^{--}$ & $2$ & $-2$\\\hline
\end{tabular}
\end{center}
Furthermore, note the important relations
\begin{eqnarray}
Q_{\text{BRST}}J^{--}&=&\tilde{G}^-,\\
Q_{\text{BRST}}\tilde{G}^+&=&J^{++}.
\end{eqnarray}
\subsubsection{Topological Amplitudes}\label{sec:topamp6d}
Now we are ready to discuss possible candidates for topological correlation functions. The first (and probably simplest) guess is 
\begin{align}
\int_{\mathcal{M}_g}\langle|\prod_{a=1}^{3g-3}G_{K3}^-(\mu_a)|^2\left[\int |\tilde{G}_{K3}^+|^2\right]^{g-1}\rangle.
\end{align}
Although the charge deficit is perfectly balanced, this expression can be easily shown to vanish identically: rewriting one of the $\tilde{G}^+$ as
\begin{align}
\tilde{G}^+=\oint \tilde{G}^+J\, ,
\label{defcon}
\end{align}
one can deform the contour of integration, such that it encircles other operators of the correlation function. The only two possibilities are however
\begin{align}
\oint \tilde{G}^+\tilde{G}^+=\oint \tilde{G}^+G^-=0.
\end{align}
Thus, we have to add another insertion, which has a non-vanishing residue with $\tilde{G}^+$ and zero charge, i.e. $J$:
\begin{align}
\mathcal{F}^{(\text{6d})}_{g}=\int_{\mathcal{M}_g}\langle&|\prod_{a=1}^{3g-3}G_{K3}^-(\mu_a)|^2\left[\int |\tilde{G}_{K3}^+|^2\right]^{g-1}\cdot\nonumber\\
&\cdot\int |J|^2\rangle.\label{top1}
\end{align}

In some cases it is useful to rewrite this result in a somewhat different form: Expressing one of the $G^-$ (say the one at $\mu_{3g-3}$) as
\begin{align}
G^-=\oint \tilde{G}^+J^{--},
\end{align}
one can again deform the contour integral around the other operators. The only possibility to get a non-vanishing answer is to encircle $J$, yielding $\tilde{G}^+$ according to (\ref{defcon}).
One then finds the alternative writing for $\mathcal{F}_g$:
\begin{align}
\mathcal{F}^{(\text{6d})}_{g}=\int_{\mathcal{M}_g}\langle&|\prod_{a=1}^{3g-4}G_{K3}^-(\mu_a)J^{--}(\mu_{3g-3})|^2\cdot\nonumber\\
&\cdot\left[\int |\tilde{G}_{K3}^+|^2\right]^{g}\rangle.\label{top2}
\end{align}
\subsubsection{Physical Couplings}
The next question is to find a possible connection between some physical string amplitudes and the topological correlation function (\ref{top1}) and (\ref{top2}). The answer to this question was given in \cite{BerVafa,conjectures}, namely the relevant couplings on the string side involve four 6d gravitons and $4g-4$ graviphotons at genus $g$:
\begin{align}
\langle R_+^4T_+^{4g-4}\rangle_g
\label{6dcoupling}
\end{align}
with $T^{ij}_{+,\mu\nu}$ the (self-dual) graviphoton field strength, where the graviphotons are left $\times $ right handed chiral spinors and $R_{+,\mu\nu\rho\tau}$ the 6d (self-dual) Riemann tensor. Although self-duality in 6 dimensions is meaningless, one can define a similar notion by contracting with the 6d Lorentz generators in the chiral representation\footnote{These are $4\times 4$ submatrices of the full $8\times 8$ Lorentz generators.}
\begin{align}
M_{\mu\nu}\to {M_{a}}^b=M_{\mu\nu}{(\sigma^{\mu\nu})_a}^b.
\end{align}
These couplings correspond to a chiral $\mathcal{F}$-term in the 6d $\mathcal{N}=2$ supergravity,
where the relevant Weyl superfield is given by
\begin{align}
({W_a}^b)^{ij}=&{(\sigma^{\mu\nu})_a}^bT^{ij}_{\mu\nu}+\nonumber\\
&+{(\sigma^{\mu\nu})_a}^{b}(\theta^i_L\sigma^{\rho\tau}\theta^j_R)R_{\mu\nu\rho\tau}+\ldots,
\end{align}
The coupling proposed in \cite{BerVafa} is given by
\begin{align}
\int d^4\theta_L\int d^4\theta_R&({W_{a_1}}^{b_1}{W_{a_2}}^{b_2}{W_{a_3}}^{b_3}{W_{a_4}}^{b_4}\cdot\nonumber\\
&\cdot\epsilon^{a_1a_2a_3a_4}\epsilon_{b_1b_2b_3b_4})^{g}.
\end{align}
\subsubsection{Duality Mapping}
As already mentioned in Section~\ref{topamp}, there  is a strong-weak coupling duality between type IIA and heterotic strings in 6 dimensions:
\begin{align}
g_s^{\text{IIA}}=\frac{1}{g_s^{\text{HET}}}.
\end{align}
Furthermore, the $\sigma$-model metric scales with the second power of the string coupling
\begin{align}
G^{\text{IIA}}=\frac{G^{\text{HET}}}{(g_s^{\text{HET}})^2}.
\end{align}
Thus, the coupling (\ref{6dcoupling}) is mapped under duality to
\begin{align}
(g_s^{\text{IIA}})^{6g-6}&\sqrt{G^{\text{IIA}}}R^4T^{4g-4}=\nonumber\\
&=(g_s^{\text{HET}})^{2g}\sqrt{G^{\text{HET}}}R^4T^{4g-4},
\end{align}
which is basically established by counting the metric factors in a similar way as in Section \ref{sec:nonren}. Hence, we see that a genus $g$ amplitude on the type IIA side is mapped to a genus $g+1$ loop amplitude in the heterotic theory.

This result raises the following question: Since the regularization procedure in Section \ref{sec:topamp6d} is not unique (recall that we just considered one specific insertion of additional operators), are there other $\mathcal{N}=4$ topological amplitudes with similar properties as in $\mathcal{N}=2$, namely, that they map to a 1-loop integral on the heterotic side?
\subsection{Compactification on $K3\times T^2$}
In order to answer this question, we change the space-time dimensionality by compactifying two additional directions on a torus. The resulting manifold $K3\times T^2$ has $\hat{c}=3$, so that the background charge deficit now matches precisely.
We can write again candidates for topological amplitudes. The first guess is \cite{AHN}
\begin{align}
\mathcal{F}_g^{(1)}=\int_{\mathcal{M}_g}&\langle |\prod_{a=1}^{g-1}G_{T^2}^-(\mu_{a})\prod_{b=g}^{3g-3}G_{K3}^-(\mu_b)\cdot\nonumber\\
&\cdot\int |J_{T^2}|^2\int |J_{K3}|^2 \rangle.
\label{Fg1}
\end{align}
In the same way as (\ref{top2}) can be derived from (\ref{top1}), one can rewrite (\ref{Fg1}) in the form
\begin{align}
\mathcal{F}_g^{(1)}=\int_{\mathcal{M}_g}&\langle |\prod_{a=1}^{g-1}G_{T^2}^-(\mu_{a})\prod_{b=g}^{3g-3}G_{K3}^-(\mu_{b})|\cdot\nonumber\\
&\cdot|J^{--}(\mu_{3g-3})|^2\int |J_{T^2}|^2 \rangle.
\end{align}

The physical amplitude which is computed by this correlation function involves two self-dual and two anti-self-dual 4d Riemann tensors, as well as $2g-2$ self-dual graviphotons:
\begin{align}
\langle R_+^2R_-^2T^{2g-2}_+\rangle_{g}.
\end{align}
Performing as before the duality map, we find
\begin{align}
(g_s^{\text{IIA}})^{4g-4}&\sqrt{G^{\text{IIA}}}R_+^2R_-^2T_+^{2g-2}=\nonumber\\
&=(g_s^{\text{HET}})^{2}\sqrt{G^{\text{HET}}}R_+^2R_-^2T_+^{2g-2}
\end{align}
implying a two loop integral representation on the heterotic side, which is still rather difficult to compute. The question then remains, if one can still do better and find an expression which maps to a 1-loop integration.

The modifications which realize this goal are to replace the two anti-self-dual gravitons by (second) derivatives of the graviscalar partner of the graviton in the $\mathcal{N}=4$ gravity multiplet, and to shift the loop order on the type II side from $g$ to $g+1$ \cite{AHN}:
\begin{align}
\langle R_+^2(\partial\partial \phi_+)^2T_+^{2g-2}\rangle_{g+1}.
\end{align}
The topological correlation function corresponding to this string amplitude reads:
\begin{align}
\mathcal{F}_g^{(3)}\equiv\int_{\mathcal{M}_g}&\langle |\prod_{a=1}^gG_{T^2}^-(\mu_{a})\prod_{b={g+1}}^{3g}G_{K3}^-(\mu_{b})\cdot\nonumber\\
&\cdot\int |J_{K3}|^2|\psi_3(w)|^2 \rangle,\label{fg3}
\end{align}
where $\psi_3$ stands for the (complex) fermionic coordinate of the torus, whose sole purpose is to soak the remaining zero modes. This fermion appears in the $\mathcal{N}=2$ $U(1)$ current of $T^2$: 
\begin{align}
J=\psi_3\bar{\psi}_3\, .
\end{align} 
After the topological twist, it has conformal dimension zero: 
$\text{dim}[\psi]=0$, 
$\text{dim}[\bar{\psi}]=1$.
Expression (\ref{fg3}) can be rewritten in the form
\begin{align}
\mathcal{F}_g^{(3)}=&\int_{\mathcal{M}_g}\langle |\prod_{a=1}^{g}|G_{T^2}^-(\mu_a)\prod_{b=g+1}^{3g-1}G_{K3}^-(\mu_{b})|^2\cdot\nonumber\\
&\cdot|J^{--}(\mu_{3g})|^2\int |J_{K3}|^2\int |\tilde{G}^+_{K3}|^2|\psi_3|^2 \rangle.
\end{align}

Performing again the duality map to the heterotic side, one finds:
\begin{align}
(g_s^{\text{IIA}})^{4g-2}&\sqrt{G^{\text{IIA}}}R_+^2(\partial\partial\phi_+)^2T_+^{2g-2}=\nonumber\\
&=\sqrt{G^{\text{HET}}}R_+^2(\partial\partial\phi_+)^2T_+^{2g-2},
\end{align}
yielding finally a 1-loop heterotic integral.
\subsection{Harmonicity Relation}
\subsubsection{Harmonicity in 6 Dimensions}
In Section \ref{N4SCFalg6D} we have considered a special superconformal algebra by picking a fixed $\mathcal{N}=2$ subalgebra inside $\mathcal{N}=4$. As was exploited in \cite{BerVafa}, this choice however is not unique, but there is a freedom of different choices, related by $SU(2)$ transformations. This means in particular that we are free to do the following redefinitions
\begin{align}
&\widehat{\tilde{G}^+}=u^L_1\tilde{G}^++u^L_2G^+,&&\!\!\widehat{G^-}=u^L_1\tilde{G}^--u^L_2\tilde{G}^-,\nonumber\\
&\widehat{\tilde{G}^-}=\bar{u}^L_2\tilde{G}^--\bar{u}^L_1G^-,&&\!\!\widehat{G^+}=\bar{u}^L_2G^++\bar{u}^L_1\tilde{G}^+,\nonumber
\end{align}
where $u^L_{1,2}$ are parameters obeying the normalization $|u^L_1|^2+|u^L_2|^2=1$. In fact, together with their counterparts from the right movers, $u_{1,2}^R$, they can be interpreted as coordinates in a $\mathcal{N}=2$ harmonic superspace, associated to the coset $\frac{SU_L(2)}{U_L(1)}\times \frac{SU_R(2)}{U_R(1)}$. Replacing now all operators in (\ref{top1}) by their corresponding redefined (hatted) quantities, the correlation function becomes a polynomial in $u^{L}_{1/2}$ and $u^{R}_{1/2}$ (for simplicity, we only display the left moving part):
\begin{align}
&F_g(u^L_1,u^L_2)=\nonumber\\
&=\!\!\!\!\sum_{n=-2g+2}^{2g-2}\!\!\frac{(4g-4)!(u_1^L)^{2g-2+n}(u_2^L)^{2g-2-n}}{(2g-2+n)!(2g-2-n)!}F_g^n.
\end{align}

Moreover, adding the following perturbation to the action
\begin{align}
&\bar{\lambda}^\alpha_{22}\oint G^+\oint \bar{G}^+\bar{\phi}_\alpha+\bar{\lambda}^\alpha_{12}\oint \tilde{G}^+\oint \bar{\tilde{G}}^+\bar{\phi}_\alpha+\nonumber\\
&+\bar{\lambda}^\alpha_{21}\oint G^+\oint \bar{\tilde{G}}^+\bar{\phi}_\alpha+\bar{\lambda}^\alpha_{11}\oint \tilde{G}^+\oint \bar{G}^+\bar{\phi}_\alpha.
\end{align}
it was shown that the following two equations hold~\cite{Ooguri:1995cp}
\begin{align}
&\sum_{i,j,k=1}^2u^R_k\epsilon_{ij}\frac{d}{du^L_i}\frac{D}{D\bar{\lambda}_{jk}^\alpha}\mathcal{F}_g^{(\text{6d})}=0,\label{trueharmL}\\
&\sum_{i,j,k=1}^2u^L_k\epsilon_{ij}\frac{d}{du^R_i}\frac{D}{D\bar{\lambda}_{kj}^\alpha}\mathcal{F}_g^{(\text{6d})}=0,\label{trueharmR}
\end{align}
generalizing $\mathcal{N}=2$ holomorphicity. Remarkably, these equations are exact, in the sense that there is no anomaly (or more importantly contact terms) modifying them. This is an important difference from the $\mathcal{N}=2$ holomorphicity relation (\ref{holoanom}).\footnote{In fact, as was pointed out in \cite{Ooguri:1995cp}, there are stronger relations
$$\epsilon_{ik}\frac{\partial}{\partial u_i^{L}}\frac{D}{D\bar{\lambda}_{kj}}\mathcal{F}_g^{(\text{6d})}=0,$$
which are however spoiled by the presence of contact terms, appearing in addition to an anomaly (boundary contributions).}
\subsection{Harmonicity in 4 Dimensions}
Generalizing the harmonicity relation for the topological $\mathcal{N}=4$ amplitudes on $K3\times T^2$, one encounters an extra difficulty due to the presence of Ramond-Ramond fields for the realization of the full $SU(4)$ $R$-symmetry of the superconformal algebra. To avoid this problem, one can make use of the heterotic dual representation of the topological amplitude $\mathcal{F}_g^{(3)}$ in (\ref{fg3}), in terms of a 1-loop torus integration. As was shown in \cite{AHN}, this amplitude is given by
\begin{align}
&\mathcal{F}_g^{(3,\text{HET})}=\int \frac{d^2\tau}{\tau_2^3}\frac{1}{\bar{\eta}^{24}} \tau_2^{2g+2} G_{g+1} \cdot\nonumber\\
&\cdot\sum_{(P_L,P_R) \in \Gamma^{(6,22)}}
\left(P^A_L\right)^{2g-2}
q^{\frac{1}{2}P_L^2}\bar{q}^{\frac{1}{2}P_R^2},\label{hetampres}
\end{align}
where $\tau=\tau_1+i\tau_2$ is the Teichm\"uller parameter of the world-sheet torus with $q=e^{2\pi i\tau}$, 
and $P_L,P_R$ are the left and right momenta of the $\Gamma^{(6,22)}$ compactification lattice. $\eta$ is the Dedekind-eta function, and $G_{n}$ is the $n$-th term in the $\lambda$-expansion of the following generating functional $G(\lambda,\tau,\bar{\tau})$, which was computed in \cite{AGNThet}
\begin{equation}
G(\lambda,\tau,\bar{\tau})=\left(\frac{2\pi i\lambda\bar{\eta}^3}{\bar{\Theta}_1
(\lambda,\bar{\tau})}\right)^2
\text{exp}\left(-\frac{\pi\lambda^2}{\tau_2}\right),
\end{equation}
where $\Theta_1$ is the usual odd theta-function.

In order to determine the harmonicity relation, we first introduce harmonic superspace variables $u_i^I$, $I=1,..,4$ (together with their conjugates $u_I^i$) parameterizing the coset manifold
\begin{align}
\frac{G}{H} \equiv \frac{SU(4)}{U(1)\times U(1)\times U(1)}.
\end{align}
Each of the $u^I$ transforms in the fundamental representation of $SU(4)$ and they differ only by their charges with respect to the three $U(1)$'s. We then pick up two of them (say $u_1$ and $u_2$) to project the graviphoton vertices of the amplitude:
\begin{align}
V_F(u_1,u_2)=\frac{1}{2\pi}\int d^2 z u_1^{i} u_2^{j}&(\partial X_{ij} -ip.\psi \psi_{ij})\cdot\nonumber\\
&\cdot\bar{\partial} X^{\mu} e^{ip.Z}(z,\bar{z})\, ,
\nonumber
\end{align}
where $X^\mu$ and $\psi^\mu$ are the space-time bosonic and fermionic coordinates, respectively, while $X_{ij}$ and $\psi_{ij}$ are the six internal ones. Here, we used the fact that the vector of $SO(6)$ can be expressed as antisymmetric tensor product of two $SU(4)$ fundamentals, to define for instance $X_{ij}=\sum_{\alpha =1}^6 X^\alpha (C\sigma_\alpha)_{ij}$, where $\sigma_\alpha$ are the $(4\times 4)$ part of the $SO(6)$ gamma matrices and $C$ is the charge conjugation matrix such that $(C\sigma_\alpha)^T=-C\sigma_\alpha$. Consequently, replacing all $2g-2$ graviphoton vertices $V_F$ of $\mathcal{F}^{(3,\text{HET})}_g$ with the projected ones renders the amplitude a function of $(u_1,u_2)$.

Defining furthermore the covariant derivatives $D_{ij, A}$ with respect to the moduli $t_{ij}^A$, by inserting the corresponding marginal $(1,1)$ operators $V_{ij,A}$: 
\begin{align}
V_{ij,A}=-\frac{1}{2\pi} \partial X_{ij} \bar{J}_A,
\end{align} 
where $\bar{J}_A$ are the right-moving currents, one can show the equation
\begin{align}
&\epsilon^{ijkl}\frac{\partial}{\partial u_1^i}\frac{\partial}{\partial u_2^j} D_{kl,A} \mathcal{F}^{(3,\text{HET})}_g(u_1,u_2) = \nonumber\\
&=(2g-2)(2g+1) u_1^i u_2^j D_{ij,A} \mathcal{F}^{(3,\text{HET})}_{g-1}(u_1,u_2).
\label{harmohet}
\end{align}
Since the right-hand side contains only $ \mathcal{F}^{(3,\text{HET})}_{g-1}$ (of lower genus than in the left-hand side), this term can be interpreted as an anomaly, just as in the $\mathcal{N}=2$ case. 
\section*{Acknowledgements}
This work was supported in part by the European Commission under the RTN contract MRTN-CT-2004-503369. 
The work of S.H. was supported by the Austrian Bundesministerium f\"ur Bildung, Wissenschaft und Kultur.
\appendix
\section{Vertex Operators}\label{app:vertex}
\subsection{Type II Closed Vertex Operators}
String vertex operators are classified according to their superghost charge (ghost picture). For these lectures, the relevant ghost pictures are $0$ and $-1/2$ as explained throughout the main body of the text. 

The vertex operator of the  $d$-dimensional graviton in the $0$-ghost picture is~\cite{Polchinski:1998rr}
\begin{align}
V^{(0)}(p,h)=:h_{\mu\nu}&\left(\partial X^\mu+ip\cdot \psi \psi^\mu\right)\cdot\nonumber\\
&\cdot\left(\bar{\partial} X^\nu+ip\cdot \bar{\psi} \bar{\psi}^\nu\right)e^{ip\cdot X}:,\nonumber\\
\mu,\nu=0,\ldots,d-&1
\end{align}
where $p$ is the (external) momentum and $h_{\mu\nu}$ its helicity. Furthermore, $X$ are space-time bosons and $\psi$ their fermionic counterparts. Note that the form of this operator does not dramatically change with the space-time dimension.

The graviphoton vertex in the $-1/2$-ghost picture reads
\begin{align}
&V^{(-1/2)}(p,\epsilon)=:e^{-\frac{1}{2}(\varphi+\bar{\varphi})}p_\nu\epsilon_{\mu}e^{ip\cdot X}\cdot\nonumber\\
&\left[S^a{(\sigma^{\mu\nu})_a}^b\bar{S}_b\Sigma(z,\bar{z})+S_{\dot{a}}{(\bar{\sigma}^{\mu\nu})^{\dot{a}}}_{\dot{b}}\bar{S}^{\dot{b}}\bar{\Sigma}(z,\bar{z})\right]:
\end{align}
where $S^a$ is a space-time spin field and $\Sigma$ the spin field of the internal SCFT. These quantities depend explicitly on the number of space-time dimensions and on the internal compactification space. The relevant ones used in these lectures are
\begin{itemize}
\item compactification on $CY_3$ (orbifold) or on $K3\times T^2$:
\begin{eqnarray}
S_1&=&e^{\frac{i}{2}(\phi_1+\phi_2)},\nonumber\\
S_2&=&e^{-\frac{i}{2}(\phi_1+\phi_2)},\nonumber\\
\Sigma&=&e^{\frac{i}{2}(\phi_3+\phi_4+\phi_5)}e^{\frac{i}{2}(\bar{\phi}_3+\bar{\phi}_4+\bar{\phi}_5)},\nonumber
\end{eqnarray}
\item compactification on $K3$
\begin{eqnarray}
S_1&=&e^{\frac{i}{2}(\phi_1+\phi_2+\phi_3)},\nonumber\\
S_2&=&e^{\frac{i}{2}(\phi_1-\phi_2-\phi_3)},\nonumber\\
S_3&=&e^{\frac{i}{2}(-\phi_1+\phi_2-\phi_3)},\nonumber\\
S_4&=&e^{\frac{i}{2}(-\phi_1-\phi_2+\phi_3)},\nonumber\\
\Sigma&=&e^{\frac{i}{2}(\phi_4+\phi_5)}e^{\frac{i}{2}(\bar{\phi}_4+\bar{\phi}_5)},\nonumber
\end{eqnarray}
\end{itemize}
where $\phi_I$ bosonizes the (complex) free fermion associated to the $I$-th plane, and the tilde stands for right movers.

Finally, one also needs the expression for the picture changing operators (PCO). Since in all computations in these lectures, only the internal part contributes, the relevant terms are:
\begin{align}
V_{\text{PCO}}=e^{\varphi+\bar{\varphi}}G^-\bar{G}^-.
\end{align}
\subsection{Heterotic String Vertex Operators}
The vertices of gauge fields and gauginos are similar to the type II graviton and graviphoton, except that the right moving part is replaced by a Kac-Moody current $\bar{J}^\alpha$, with $\alpha$ the index of the gauge group:
\begin{align}
&V^{(0)}(p,\varepsilon)=: \varepsilon_{\mu}\left(\partial X^\mu+ip\cdot \psi \psi^\mu\right)\bar{J}^\alpha e^{ip\cdot X}:,\nonumber\\
&V^{(-1/2)}(p)=:e^{-\frac{\varphi}{2}}e^{ip\cdot X}\cdot\nonumber\\
&\hspace{0.7cm}\cdot\left[S^a\bar{J}^\alpha_b\Sigma(z,\bar{z})+S_{\dot{a}}\bar{J}^{\alpha,\dot{b}}\bar{\Sigma}(z,\bar{z})\right]:,\nonumber\\
&V_{\text{PCO}}=e^{\varphi}G^-,\nonumber
\end{align}
where the explicit expressions of all quantities are precisely as in the type II string case.

\end{document}